\documentclass{article}

\usepackage[final]{neurips_2020}

\usepackage[utf8]{inputenc} 
\usepackage[T1]{fontenc}    
\usepackage{hyperref}       
\usepackage{url}            
\usepackage{booktabs}       
\usepackage{amsfonts}       
\usepackage{nicefrac}       
\usepackage{microtype}      

\usepackage{yk}

\usepackage{xcolor}

\title{Black Loans Matter: Distributionally Robust Fairness for Fighting Subgroup Discrimination}

\author{
  Mark Weber \\
  IBM Research \\
  MIT-IBM Watson AI Lab \\
  \texttt{mark.weber@ibm.com} \\
  
    \And
    Mikhail Yurochkin \\
    IBM Research \\
    MIT-IBM Watson AI Lab \\
    \texttt{mikhail.yurochkin@ibm.com} \\
  
    \And
    Sherif Botros \\
    Wells Fargo \\
    \texttt{sherif.m.botros@wellsfargo.com} \\

    \And
    Vanio Markov \\
    Wells Fargo \\
    \texttt{vanio.markov@wellsfargo.com} \\
}

\begin{document}

\maketitle

\begin{abstract}

Algorithmic fairness in lending today relies on group fairness metrics for monitoring statistical parity across protected groups. This approach is vulnerable to subgroup discrimination by proxy, carrying significant risks of legal and reputational damage for lenders and blatantly unfair outcomes for borrowers. Practical challenges arise from the many possible combinations and subsets of protected groups. We motivate this problem against the backdrop of historical and residual racism in the United States polluting all available training data and raising public sensitivity to algorithimic bias. We review the current regulatory compliance protocols for fairness in lending and discuss their limitations relative to the contributions state-of-the-art fairness methods may afford. We propose a solution for addressing subgroup discrimination, while adhering to existing group fairness requirements, from recent developments in individual fairness methods and corresponding fair metric learning algorithms.

\end{abstract}

\section{Introduction}

Lending is a key opportunity space for artificial intelligence in banking. It is also a domain fraught with intractable bias due to our reliance on historical data, which is irreparably poisoned by structural and cultural racism, past and present. This problem is particularly acute in the United States, which continues to grapple with its racial history and the residuals of injustice therein. Today, tens of millions of creditworthy prospective borrowers are excluded from credit access or charged with ``racial premiums'' on interest rates because their creditworthiness is determined by models failing to disambiguate historical oppression and present merit. As a result, machine learning can become a vehicle for perpetuating and even accelerating prejudice in a reinforcing loop of bias training data. 

In a landmark study of 3.2 million mortgage applications and 10.0 million refinance applications from Fannie Mae and Freddie Mac, \citet{Bartlett} found evidence of racial discrimination in face-to-face lending as well as in algorithmic lending. Overall, Latinx and African-American indviduals suffered rejection rates of greater than 60.6\% versus 47.6\% for everyone else. Moreover, Latinx and African-American borrowers payed racial premiums of 7.9 and 3.6 basis points more in interest for home-purchase and refinance mortgages, amounting to an additional \$765M per annum, a staggering disparity for modern times. Algorithmic lending was found to improve acceptance rate parity, but it was guilty of imposing racial premiums of 5.3 basis points for purchase mortgages and 2.0 basis points for refinance mortgages. This is the algorithmic equivalent of reverse redlining, the unlawful predatory practice of targeting minorities for exploitative interest rates relative to their white counterparts. The researchers speculate this may be due to the ability of an algorithm to discriminate on the basis of intragroup variation such as neighborhood or shopping behavior. The fact that such discrimination can pass the compliance protocols of lenders is the subject of this paper.

Meanwhile, the Consumer Financial Protection Bureau (CFPB) estimates 20\% of the U.S. adult population is under-served for credit and this population is more likely to be from minority and protected groups \citep{CFPB-invisible}. These are the so called ``credit invisible'' customers who are undocumented by the major credit reporting agencies, as well as the ``thin-file'' customers who are visible but whose files are either incomplete or stale for the purpose of accurately predicting creditworthiness.

The problem of algorithmic discrimination is being raised at the highest levels of government and society. U.S. senators \citet{Warren} cited \citet{Bartlett} along with other studies in a special letter to the Board of Governors of the Federal Reserve System, the Office of the Comptroller of the Currency, the Federal Deposit Insurance Corporation, and the Consumer Financial Protection Bureau, with detailed admonishment to do more to combat algorithmic lending discrimination. Amidst a resurgence of social justice activism led by Black Lives Matter in 2020, we are especially mindful of lending outcomes for black Americans, whose financial prospects have been most undermined by structural and cultural prejudice, past and present. As modern decision-making processes have become increasingly dependent on machine learning, public backlash has spiked in numerous sectors as people grapple with the incongruence of modern values and the historical data skeletons unearthed by our amoral learning algorithms.

Researchers, lending practitioners, and policymakers alike should pay heed to the zeitgeist. Predatory lending practices notwithstanding, access to affordable credit is an important tool for extending equal opportunity to minority households and minority-owned businesses; this is the public policy motivation. Banks being principally motivated by profit, expansion of the customer lending base is a business opportunity aligned with the societal interests of financial inclusion. Add the risks of government fines and reputational damage, which are important checks on predatory practices disguised as inclusion, and fairness is clearly a profitability factor in modern times. 

Banking and regulation move slowly, but the field of artificial intelligence moves fast. The purpose of this paper is to flag a key vulnerability in the standard fairness protocols in lending, namely group fairness via statistical parity, which we speculate may be a factor in the aforementioned algorithmic injustices that somehow persist in the face of rigorous compliance protocols. We evaluate the problem of subgroup discrimination and explain how recent advances in individual fairness may have a contribution to make to the cause of justice in lending.

\section{Fairness in Lending}

Access to affordable credit is an important factor in socioeconomic stability and professional outcomes for individuals and households around the world. As such, governments have enacted laws and regulations to protect consumers from unfair discrimination when applying for credit. In the United States, the two most relevant laws are the Fair Credit Reporting Act (FCRA) \citep{FCRA} and the Equal Credit Opportunity Act (ECOA) \citep{ECOA}. ECOA is the law designed to protect consumers against discrimination. It prohibits lenders from discriminating based on certain \textit{protected attributes} such as race, gender, national origin, religion, age, marital status or whether the applicant receives public assistance. There are two criteria for claiming discrimination: disparate treatment and disparate impact. \textit{Disparate treatment} is when a lender treats a consumer differently based on a protected characteristic such as race. \textit{Disparate impact} occurs when the lender’s policy or algorithm results in a disproportionate negative impact on a protected group. For the purpose of this paper, we define fairness simply as the avoidance of disparate treatment and disparate impact, but we refine this definition to account for subgroup complexity.


\subsection{Inside the bank}

There are three key aspects to the enforcement of fairness in lending inside a bank. The first involves the organizational structure of the bank, the second involves the data that can be collected and considered in modeling, and the third involves the statistical techniques used by compliance officers to monitor fairness.

\paragraph{Separation} In adherence to current regulation, banks have internal walls between model development and model risk, meaning model developers are actively excluded from the process of ensuring algorithmic fairness. This is intended to prevent model developers from gaming the system. However, this silo structure largely prevents model developers from evaluating and using state-of-the-art fairness algorithms, and it drives inefficiency for the bank; both undermine the objective of producing fair algorithms.

\paragraph{Blindness} Generally, neither model developers nor compliance officers have access to protected attributes because banks are not allowed to collect this data (mortgage applications are an exception because applicants can volunteer some protected information such as race). Since it is difficult to ensure fairness for protected groups when protected attributes are unavailable, the  Consumer Financial Protection Bureau releases guidance on how to construct proxies for race and ethnicity from public information such as census data. Specifically, the Bureau’s Office of Research (OR) and Division of Supervision, Enforcement, and Fair Lending (SEFL) use a method called Bayesian Improved Surname Geocoding (BISG) to combine geography and surname-based information into a single proxy probability for race and ethnicity \citep{CFPB-proxy}. This is a strange policy framework: we blind ourselves from protected attributes, only to reconstruct probabilistic proxies for those same protected attributes. 

\paragraph{Group Fairness} Group fairness metrics are utilized by a bank's risk office to compute some group statistic for all protected groups; these should more or less match the statistics of the majority group. These are among the most common group fairness metrics:

\begin{itemize}
    \item \textit{Demographic parity}. The distribution of outcomes for each group should be the same. For example the percentage of loans accepted from the majority group should be more or less the same as the percentage accepted from different minority groups.
    \item \textit{Equal Odds}. Conditioned on the target variable (for example people who paid their loans and people who did not), the groups should have the same statistics. For example, for all the people who paid their loans, the algorithm should have the same true positive, false positive, true negative and false negative rates for all groups. Similarly for all people who did not pay their loans. This allows for the possibility that different groups may have different rates of default. 
    \item \textit{Equal Opportunity}. This is similar to equal odds, except that we do not care about all statistics. We only care about the positive outcomes like loan approvals. For example, the acceptance rate for loans among the pool of qualified applicants should be the same for all groups. While equal odds and equal opportunities allow for more accuracy in the case when there are real discrepancies between groups, it does not fix any historical injustice. 
    \item \textit{Prediction Rate Parity}. This is another metric that is conditioned on the prediction. For any prediction value (e.g. person will likely default or not), the true and false rates for the different groups should match. 
\end{itemize}

\section{Subgroup Discrimination}

The aforementioned group fairness metrics have a limitation: they require a definition of the protected group (e.g. African Americans). In the real world, identity is much more complex and heterogeneous with numerous subgroups and subgroup combinations within and across protected groups. This has been observed by scholars of behavioral medicine, for example, who argue that inattention to subgroup diversity can result in disparate outcomes for significant minority populations \citep{Birnbaum-Weitzman2013}. Group-level statistical metrics can easily be manipulated by humans and (especially) by machine learning algorithms enforcing fairness as a constrained optimization problem. A given subgroup may be large enough to pose a legal and reputational risk if treated unfairly while remaining sufficiently small enough to pass the meta-group fairness metrics. We call this subgroup discrimination.


Suppose a machine learning model optimizing for accuracy within group fairness constraints decides to use zipcode proxies for race to favor African Americans in the wealthiest black communities in America (see ``Black Type 1'' in Table \ref{table:zip}) with extraordinarily favorable credit terms (more favorable than white customers of the same financial means) while offering extremely unfavorable terms to African Americans everywhere else (more unfavorable than their white counterparts). The hypothetical values in the table show how group parity can be satisfied even while subgroup discrimination takes place. Similar individuals are clearly not treated similarly irrespective of race. Not only is this problematic for the cause of true fairness, such group fairness policies also pose serious regulatory and reputational risks to institutions relying on them.

Consider the 2019 controversy over the Goldman Sachs credit card collaboration with Apple, which triggered a now pending government investigation. Viral anecdotal accounts from Twitter indicated that Goldman Sachs' credit limit policy discriminated against women. Reviewing the details of the claims and Goldman Sachs' response, we speculate the policy may have penalized homemakers relative to their working partners by virtue of prioritizing personal income and employment status while ignoring household-level financials. The majority of homemakers in the United States today are female. Thus, a significant subgroup of women may indeed have received disparate treatment, even while all women as a protected group may have been treated with statistical parity to men, which we may safely assume based on the fact that the policy passed Goldman Sachs' model risk assessments before release. This is illustrative of our argument that it is not only the whole group that poses risk, but sufficiently large subgroups (e.g. homemakers) within a protected meta-group.

\begin{table}
\caption{Simple example of subgroup discrimination, where group fairness by statistical parity is still satisfied, and where Type 1's and Type 2's respectively are similar to one another except for race.}
\vskip -0.1in
\label{table:zip}
\begin{center}
\begin{tabular}{lccccccc}
\toprule
{} & Black Type 1 & Black Type 2 & White Type 1 & White Type 2 \\
\midrule
Acceptance Rate & 80\% & 20\% & 60\% & 40\% \\
\midrule
Annual Percentage Rate (APR) & 10\% & 30\% & 25\% & 15\% \\
\bottomrule
\end{tabular}
\end{center}
\end{table}

\section{Algorithms for Equal Treatment Across Subgroups}

Subgroup fairness has been proposed in the literature as a way to interpolate between group and individual fairness \citep{hebert2017calibration,kearns2018preventing,kim2019multiaccuracy}. When there is a limited number of known subgroups, it can be achieved with group fairness algorithms. However when subgroups are not clearly defined, or there are many of them, individual fairness is a more practical approach.

In banking, we could consider enforcing group fairness not only for a limited set of protected attributes, e.g. race and gender, but also their intersections. For example, achieving statistical parity on four groups simultaneously: African-American women, Africa-American men, White women and White men. This problem can be solved with group fairness techniques. We note that such approach, while helpful to improve fairness overall, is unlikely to prevent the zip code example discrimination as it does not consider low income zip code as a subgroup. Adding ``zip code'' as a protected group could prevent the problem, but it is practically difficult if not impossible to anticipate all possible subgroups and combinations of subgroups. This problem becomes more pronounced with the increasing number of a features being utilized by a machine learning model in modern banking, which is optimizing for profit within risk constraints, and may prove quite sophisticated at exotic forms of subgroup discrimination.

Individual fairness emerges as a mechanism by which we can deal with subgroup complexity. We will outline our algorithmic recommendations for machine learning practitioners in banking. We will also provide a real data example where enforcing individual fairness prevented an unaccounted latent discrimination, while pursuing group fairness actually exacerbated it.

\subsection{Individual fairness}
We review individual fairness definitions and the recent algorithmic advances for achieving it in practice.

Let a machine learning model be a map $h:\cX\to\cY$, where $\cX$ and $\cY$ are the input and output spaces. \citet{dwork2011Fairness} defined individual fairness as follows:
\begin{equation}
d_y(h(x_1),h(x_2)) \le Ld_x(x_1,x_2)\text{ for all }x_1,x_2\in\cX,
\label{eq:metricFairness}
\end{equation}
where $d_x$ and $d_y$ are metrics on the input and output spaces and $L\ge 0$ is a Lipschitz constant. The fair metric $d_x$ encodes our intuition of which samples should be treated similarly by the machine learning model. There are two major challenges: (i) it is not possible to provide any practical generalization guarantees due to a requirement that the inequality holds for all possible pairs of individuals; (ii) defining an appropriate fair metric.

\citet{Yurochkin2020Training} proposed Distributionally Robust Fairness (DRF), a distributional definition of individual fairness to address the first challenge. They define a machine learning model $h:\cX\to\cY$ as $(\eps,\delta)$-DRF with respect to the fair metric $d_x$ if and only if
\begin{equation}
\max\limits_{P:W(P,P_n) \le \eps} \int_{\cX \times \cY}\ell(x,y,h)dP(x,y) \le \delta,
\label{eq:correspondenceFairness}
\end{equation}
where $\ell:\cX \times \cY \times\cH\to\reals_+$ is a loss function, $h \in \cH$ is the machine learning model, $P_n$ is the empirical distribution of the data, $\eps > 0$ is a small tolerance parameter, and $W(P,P_n)$ is the Wasserstein distance between distributions $P$ and $P_n$ with the cost $d((x_1,y_1),(x_2,y_2)) = d_x(x_1,x_2) + \infty \cdot \mathbf{1}(y_1 \neq y_2)$. Comparing this definition to individual fairness in \eqref{eq:metricFairness}, here the loss function replaces the metric on the output space $d_y$ and the tolerance parameter $\eps$ is analogous to the Lipshitz constant $L$. \textbf{The crucial difference is that DRF \eqref{eq:correspondenceFairness} considers average violations of individual fairness making it amendable to statistical analysis establishing generalization guarantees.} DRF measures by how much the loss can increase when considering the fair neighborhood of the observed data. 

In the neighborhood redlining and reverse-redlining example, if the fair metric $d_x$ considers individuals with similar household income to be similar regardless of their zip code, where zip code can be used as a proxy for race, DRF will raise an alarm by taking $P$ as $P_n$ where the zip codes of all with high household income is set to the zip codes of those with the low income records a significant loss increase.

\citet{Yurochkin2020Training} derived the dual formulation of \eqref{eq:correspondenceFairness} amendable to an efficient training algorithm, called Sensitive Subspace Robustness (SenSR), similar to those used in the adversarial robustness literature \citep{madry2017towards,sinha2017certifying}.

\subsection{Fair metrics}
In general, individual fairness does not imply group fairness, which may seem unsuitable in the context of existing legal regulations in finance. It may also seem that defining an appropriate fair metric requires significant expertise, such as knowing which zip codes are more or less racially homogeneous. Fortunately, recent literature offers a variety of ways to learn fair metric from data, \citep{ilvento2019Metric,mukherjee2020Two,Yurochkin2020Training,wang2019Empirical} alleviating the need for the expert knowledge. A subset of these practises can account for the information about protected groups, bridging the gap between individual and group fairness to satisfy existing regulations. We review two approaches.

First, consider a Mahalanobis-type fair metric of the form:
\begin{equation}
\label{eq:metric}
d_x(x_1, x_2) = \langle (x_1 - x_2), \Sigma (x_1 - x_2) \rangle,
\end{equation}
where $\Sigma$ is a positive semi-definite matrix. $\Sigma$ can control what features should be ignored or amplified when evaluating similarity between individuals. \citet{Yurochkin2020Training} proposed to construct $\Sigma$ based on the \emph{sensitive subspace}. This subspace contains variation in the data due to protected attributes, such as differences between black and white individuals. Let $V = \{v_1,\dots,v_k\}$ be a set of \emph{sensitive directions}, i.e. vectors spanning a sensitive subspace. Then $\Sigma = I - P_V$, where $P_V$ is the projection onto the span of $V$ ($\Sigma$ is an orthocomplement projection). Sensitive directions can be obtained by fitting linear predictors to sensitive attributes and taking vectors orthogonal to the decision boundary. For example, let $v$ be a vector of coefficients of a logistic regression fitted to predict race (black or white here). Then $v$ is a sensitive direction. An individually fair classifier trained with such a fair metric will treat individuals of different races similarly while correctly ignoring features in the data that correlate with race (i.e. racial proxies). Because this fair metric accounts for the protected attribute, intuitively, the corresponding individually fair classifier should also satisfy group fairness, which is important for satisfying existing legal regulations. It is then straightforward to account for multiple protected groups as well as their subsets and intersections (i.e. subgroups).

The second approach we review also utilizes Mahalanobis-type fair metric as in \eqref{eq:metric}, but uses a different procedure to learn $\Sigma$. \citet{mukherjee2020Two} proposed EXPLORE, an algorithm to learn $\Sigma$ from pairs of comparable and incomparable samples. On a high level, EXPLORE is a logistic regression variant on input pairs. It learns $\Sigma$ such that when fair distance is small, the corresponding logistic regression predicts ``comparable,'' and when the fair distance is large it predicts ``incomparable.'' Consider race as protected attribute. To bridge the gap between individual and group fairness, they defined pairs for learning $\Sigma$ as follows: any pair of individuals with same class label but different race are comparable; any pair of individuals with different class label are incomparable.

\begin{table}
\caption{Summary of the \texttt{Adult} experiment from \citet{mukherjee2020Two}.}
\vskip -0.1in
\label{table:adult}
\begin{center}
\begin{tabular}{lccccccc}
\toprule
{} &        B-Acc,\% &   $\text{S-Con.}$ &       $\mathrm{Gap}_G^{\mathrm{RMS}}$ &       $\mathrm{Gap}_R^{\mathrm{RMS}}$ &     $\mathrm{Gap}_G^{\mathrm{max}}$ & $\mathrm{Gap}_R^{\mathrm{max}}$ \\
\midrule
SenSR+EXPLORE & 79.4 & \textbf{.966} & \textbf{.065} & \textbf{.044} & \textbf{.084} & \textbf{.059} \\
SenSR &  78.9 & .934 & .068 & .055 & .087 & .067 \\
Baseline &  \textbf{82.9} & .848 & .179 & .089 & .216 & .105 \\
Adv. debiasing &  81.5 & .807 & .082 & .070 & .110 & .078  \\
\bottomrule
\end{tabular}
\end{center}
\end{table}

\subsection{A case study: income prediction from Census data}

To illustrate how individual fairness can prevent subgroup discrimination, including subgroups not explicitly considered at training time and while satisfying legal group fairness regulations, we review the experiment considered in \citet{Yurochkin2020Training,mukherjee2020Two}. The task is to predict individuals' income based on the Census data, known as the \texttt{Adult} dataset \citep{bache2013UCI} in the fairness literature. Specifically, we discuss the performance of SenSR trained with two fair metrics discussed in the previous section, naive training by empirical risk minimization, and one of the prior group fairness algorithms.

The results of their experiment are summarized in Table \ref{table:adult}. First we describe the comparison metrics: B-Acc stands for balanced accuracy measuring classification performance; Gap metrics measure equal odds group fairness violations (0 is an ideal value); S-Con. stands for spouse consistency and measures the counterfactual invariance of the classifier's prediction when the ``relationship'' feature is modified. The ``relationship'' feature takes values such as ``husband',' ``wife,'' and ``unmarried.'' This is an example of a feature that correlates with one of the protected attributes (gender), but could be easily overlooked at the training stage (analogous to the ``zip code'' feature in our earlier example). In this experiment ``relationship'' was treated as any other general (i.e. not protected) feature and thus may become the basis of an overlooked subgroup.

Our first observation, is that enforcing group fairness may result in a subgroup of women, e.g. unmarried women, being treated unfairly even when the equal odds violation for gender is small. Adversarial debiasing \citep{zhang2018Mitigating} is one of the recent group fairness methods at our disposal: it reduces group fairness violations in comparison to the Baseline (i.e. naive training). However, it actually \emph{worsens} spouse consistency, producing a significant subgroup fairness violation. In contrast, enforcing individual fairness with SenSR (either using sensitive subspace or EXPLORE to obtain fair metric from data) yields a higher spouse-consistency score \textit{and} the least amount of group fairness violations. This experiment illustrates that individual fairness is effective for satisfying legal group fairness requirements, as well as safeguarding against unintentional subgroup discrimination.

\section{Recommendations}

Having reviewed the subgroup discrimination vulnerability of group fairness and the improvement opportunities individual fairness algorithms afford, we return to the specifics of lending and the practical considerations for implementing fair algorithms.


\subsection{For lenders}

Training an individually fair classifier consists of three stages: (i) defining/learning a fair metric; (ii) training an individually fair model with respect to the given fair metric; (iii) assessing fairness and performance. The aforementioned departmental separation of model development and model risk in banks naturally aligns with this pipeline. We recommend the risk department take charge of (i) and (iii) and the modeling department take charge of (ii). While defining and learning the fair metric may require protected attributes (or proxies) to promote group fairness, the fair metric itself does not explicitly expose protected attributes (or their proxies). This observation suggests that it may be permissible for a risk department to provide a fair metric for the modeling department allowing undue gamesmanship of the metrics.

On the other hand, the department separation goes against modern group fairness algorithms. For example, \citet{agarwal2018Reductions} proposed to formulate group fairness as a constraint optimization problem. The constraints are defined based on the group fairness requirements and corresponding protected attributes (or their proxies) and need to be evaluated routinely during training. In the context of banking, the model risk department defines group fairness requirements, hence the constraints, but keeps these and the protected attributes (or proxies) secret from the modelers, making it impossible to efficiently solve the corresponding constraint optimization problem during training. This undermines the cause of fairness and further motivates the adoption of our recommendations for a shift to individual fairness with the departmental delineations described above.

\subsection{For policymakers}

In a feature rich environment, machine learning models can learn a variety of sophisticated subgroup discrimination schemes while satisfying group fairness metrics for compliance. We therefore urge policymakers to update the regulatory guidances to account for subgroup discrimination and the corresponding algorithmic toolbox available to lenders. Banks should be encouraged if not required to use state-of-the-art methods for algorithmic fairness.

We also recommend a re-evaluation of the requirement to blind model developers from sensitive attributes. While we provide an approach for lenders that satisfies this constraint, the practice both limits the algorithmic fairness toolbox for banks and drives inefficiency in the development of certifiably fair models, which ultimately affects the people seeking fair treatment at any given moment in time. The approach seems akin to the now debased ideology of ''color blindness,'' criticized by sociologists and justice advocates as promoting willfull ignorance of and complacence with structural and cultural systems of racial oppression. In other words, pretending race does not exist does not result in fair outcomes; in fact, the opposite happens because the underlying problem of racial injustice is not properly dealt with or accounted for. Similarly, ``fairness by unawareness'' in lending is susceptible to racial discrimination because model developers can hide behind a guise of objectivity while a machine learning model learns how to discriminate by proxy and/or by subgroup. This topic requires further examination, but we recommend for consideration the voluntary inclusion of one's racial identification information in the service of improving algorithmic fairness, where banks are disallowed from claiming racial ignorance i.e. ``color blindness.'' Since the government is already providing sophisticated racial proxies, of which the general citizenry is likely unaware, it seems appropriate to consider abandoning this practice in favor of a more transparent, voluntary approach that may also yield more just outcomes for minorities.


\section{The urgency of algorithmic fairness}

We have observed that both traditional lending and algorithmic lending are subject to racial discrimination. We have reviewed the current banking practices on fairness in lending. We have flagged the problem of subgroup discrimination and discussed methods to mitigate it. We have recommended individual fairness frameworks including distributionally robust fairness (DRF) and we have explained how they can be implemented within the current compliance regime. We have challenged the notion that separation and blindness are effective in promoting algorithmic fairness in lending.

The human tapestry is not as black-and-white as black and white. It is rich in complex subgroups and combinations. No algorithm can change the history of racial oppression in the United States, nor male privilege, nor any injustice embedded in historical data. Nevertheless, we have a responsibility to find ways to to make sure that each and every person is treated fairly, not just special subsets of people, even within a protected class. Decisions made on the basis of racially bias historical data produce racially bias results, which in turn become the data we train on tomorrow. This threatens to perpetuate and accelerate racial bias by propagating it into the future. This why fairness in lending, and indeed in every part of our increasingly algorithmic society, is not only an important issue, but an urgent one.


\section*{Broader Impact}

This work addresses the ethical and societal consequences of suboptimal policies and practices in algorithmic fairness for lending. If correct in its analysis, our work should benefit groups and subgroups who suffer from prejudiced historical data affecting current lending decisions by virtue of enabling banks to identify good prospective borrowers irrespective of protected attributes and proxies. It should also benefit lenders who seek to avoid legal and reputational risk by violating, or being perceived to violate, fairness regulations and public principles in lending. This topic is extremely complex, particularly given the racial history of the United States. Our recommendations are measured accordingly, not as definitive solutions but as prudent considerations for model developers, compliance officers, and policy makers charged with ensuring fairness. Such experts should evaluate our arguments and weigh the potential problems or unintended consequences of our recommendations.




\small
\bibliographystyle{plainnat}
\bibliography{bibliography}

\end{document}